\documentclass[aps,showpacs,prd,twocolumn]{revtex4}

\usepackage{amssymb}
\usepackage{amsfonts}
\usepackage{amsmath}
\usepackage{graphicx}
\usepackage{bm}
\usepackage{color}
\usepackage[dvips]{epsfig}
\usepackage[dvips]{graphicx}
\usepackage{float}
\usepackage{epstopdf}
\usepackage[colorlinks=true,pdfstartview=FitV,linkcolor=blue,citecolor=blue,urlcolor=blue,breaklinks=true]{hyperref}

\begin{document}

\title{Self-dual configurations in a generalized Abelian Chern-Simons-Higgs
model with explicit breaking of the Lorentz covariance}
\author{Rodolfo Casana}
\email{rodolfo.casana@gmail.com}
\affiliation{Departamento de F\'\i sica, Universidade Federal do Maranh\~ao, 65080-805,
S\~ao Lu\'is, Maranh\~ao, Brazil}
\author{Lucas Sourrouille}
\email{lsourrouille@yahoo.es}
\affiliation{Universidad Nacional Arturo Jauretche, 1888, Florencio Varela, Buenos Aires,
Argentina}

\begin{abstract}
\centerline{\textbf{Abstract}} We  have studied  the existence of
self-dual solitonic solutions in a generalization of the Abelian
Chern-Simons-Higgs model. Such a generalization introduces two different
nonnegative functions, $\omega_1(|\phi|)$ and $\omega(|\phi|)$, which split
the  kinetic term of the Higgs field - $|D_\mu\phi|^2 \rightarrow
\omega_1 (|\phi|)|D_0\phi|^2-\omega(|\phi|) |D_k\phi|^2$ - breaking
explicitly  the Lorentz covariance. We have shown that a clean
implementation of the Bogomolnyi procedure only can be implemented whether $%
\omega(|\phi|) \propto \beta |\phi|^{2\beta-2}$ with $\beta\geq 1$. The
self-dual or Bogomolnyi equations produce an infinity number of soliton
solutions by choosing conveniently the generalizing function $%
\omega_1(|\phi|)$ which must be able to provide a finite magnetic field.
Also, we have shown that by properly choosing the generalizing
functions it is possible to reproduce the Bogomolnyi equations of  the
Abelian Maxwell-Higgs and Chern-Simons-Higgs models. Finally, some new
self-dual $|\phi|^6$-vortex solutions have been analyzed both from
theoretical and numerical point of view.
\end{abstract}

\maketitle



\section{Introduction}

A time ago it was shown that (1+2)-dimensional matter field interacting with
gauge fields whose dynamics is governed by a Chern-Simons term support
soliton solutions \cite{Jk1}, \cite{Jk2} (for a review see Refs. \cite{hor1}%
, \cite{hor2}, \cite{hor3}, \cite{hor4} and \cite{hor5}). These models have
the particularity to become self-dual when the self-interactions are
suitably chosen \cite{JW1}, \cite{JW2}, \cite{JP1}, \cite{JP2}. When
self-duality occurs the model presents interesting mathematical and physical
properties, such as the second order Euler-Lagrange equations can be solved
by a set of first-order differential equations \cite{Bogo}, \cite{Vega} and,
the model admits a supersymmetric extension \cite{LLW}. The Chern-Simons
gauge field dynamic remains the same when coupled with matter-fields either
relativistic \cite{JW1}, \cite{JW2} or nonrelativistic \cite{JP1}, \cite{JP2}%
. In addition the nature of the soliton solutions can be topological and/or
nontopological \cite{JLW}.

The inclusion of non-linear terms to the kinetic part of the Lagrangian has
interesting consequences, as for example, the existence of topological
defects without a symmetry-breaking potential term \cite{sy}. In the recent
years, theories with nonstandard kinetic term, named $k$-field models, have
received much attention. The $k$-field models are mainly in connection with
effective cosmological models \cite{APDM1}, \cite{APDM2}, \cite{APDM3}, \cite%
{APDM4}, \cite{APDM5}, \cite{APDM6}, \cite{APDM7}, as well as the tachyon
matter \cite{12} and the ghost condensates \cite{131}, \cite{132}, \cite{133}%
, \cite{134}, \cite{135}. The strong gravitational waves \cite{MV} and dark
matter \cite{APL}, are also examples of non-canonical fields in cosmology.
The investigations concerning to the topological structure of the $k$%
-field theories have shown that they support topological soliton solutions
both in pure matter models as in gauged field models \cite{BAi1}, \cite%
{BAi2}, \cite{BAi3},\cite{BAi4}, \cite{BAi5}, \cite{BAi6}, \cite{BAi7}, \cite%
{BAi8}, \cite{SG1},\cite{SG2}, \cite{SG3}, \cite{SG4}, \cite{SG5}, \cite{SG6}%
, \cite{lucas1},\cite{lucas2}, \cite{lucas3}, \cite{lucas4}. These
solitons have certain features which are not necessarily shared with those
of the standard models \cite{B1}, \cite{B2}, \cite{B3}.

The aim of this manuscript is to study a Chern-Simons-Higgs model with a
generalized dynamics which breaks Lorentz covariance, i.e,
\begin{equation}
|D_{\mu }\phi |^{2}\rightarrow \omega _{1}(|\phi |)|D_{0}\phi |^{2}-\omega
(|\phi |)|D_{i}\phi |^{2}.
\end{equation}%
The nonstandard dynamics is introduced by the functions $\omega _{1}$ and $%
\omega $, depending on the Higgs field. During the implementation of
Bogomolnyi trick is demonstrated that self-dual configurations exist if the
function $\omega $ is proportional to $|\phi |^{2\beta -2}$ with $\beta \geq
1$. On the other hand, the function $\omega _{1}$ remains arbitrary but near
the origin should behave as $|\phi |^{2\delta }$ with $\delta \geq -1$ in
order to have a well behavior for the magnetic field. In particular
we have chosen the functions $\omega_{1}$ and $\omega $, to be
\begin{equation}
\omega _{1}(|\phi |)=|\phi |^{2M},~\omega (|\phi |)=(N+1)|\phi |^{2N},
\end{equation}%
where $M\geq -1$ and $N\geq 0$.  This way, the Bogomolnyi equations
produce an infinite number of soliton solutions, one for each value of the
pair $(N, M)$. It is possible to show that for particular values of $N$, $M$%
, the Bogomolnyi equations of the Maxwell-Higgs or Chern-Simons Higgs models
can be recuperated. Finally, we have constructed, analytically and
numerically, novel soliton solutions for some values of $N$ and $M$.

\section{The theoretical framework}

Following the same  ideas introduced in  Refs. \cite{BAi1}, \cite%
{BAi2}, \cite{BAi3}, \cite{BAi4}, \cite{BAi5}, \cite{BAi6}, \cite{BAi7},
\cite{BAi8}, \cite{SG1}, \cite{SG2}, \cite{SG3}, \cite{SG4}, \cite{SG5},
\cite{SG6}, \cite{lucas1}, \cite{lucas2}, \cite{lucas3}, \cite{lucas4}, we
start by considering a generalized $(2+1)$-dimensional Chern-Simons-Higgs
(CSH) model where the  complex scalar field possess a modified dynamic. Such a model is described by the following action,
\begin{equation}
S=S_{cs}+\!\!\int \!d^{3}x\!\left[ \!\frac{{}}{{}}\omega _{1}(|\phi
|)|D_{0}\phi |^{2}-\omega (|\phi |)|D_{i}\phi |^{2}-V(|\phi |)\right] \!\!,
\label{Acg1}
\end{equation}%
where $S_{cs}$  represents the Chern-Simons action given  by
\begin{equation}
S_{cs}=\int \!d^{3}x\,\frac{\kappa }{4}\epsilon ^{\mu \nu \rho }A_{\mu
}F_{\nu \rho }.
\end{equation}%
The covariant derivative $D_{\mu }\phi $ is defined by
\begin{equation}
D_{\mu }\phi =\partial _{\mu }\phi -ieA_{\mu }\phi ,
\end{equation}%
with $\mu =0,1,2$. The metric tensor is $g_{\mu \nu }=(1,-1,-1)$ and
$\epsilon ^{\mu \nu \rho }(\epsilon ^{012}=1)$ is the totally antisymmetric
Levi-Civita tensor.

In action (\ref{Acg1}) we notice\ the usual Higgs kinetic term, $|D_{\mu
}\phi |^{2}=|D_{0}\phi |^{2}-|D_{k}\phi |^{2}$, was replaced by a more
generalized term, $\omega _{1}(\left\vert \phi \right\vert )|D_{0}\phi
|^{2}-\omega (\left\vert \phi \right\vert )|D_{k}\phi |^{2}$, which breaks
explicitly the Lorentz covariance. The dimensionless functions $\omega
_{1}(\left\vert \phi \right\vert )$ and $\omega (\left\vert \phi \right\vert
)$ are nonnegative and, in principle, arbitrary functions of the complex
scalar field $\phi $. The function $V(\left\vert \phi \right\vert )$ is a
self-interacting scalar potential.

The gauge field equation obtained from the action (\ref{Acg1})\ is
given by
\begin{equation}
\frac{\kappa }{2}\epsilon ^{\mu \alpha \beta }F_{\alpha \beta }-e\mathcal{J}%
^{\mu }=0,  \label{ge0}
\end{equation}%
with $\mathcal{J}^{\mu }=\omega _{1}\delta _{0}^{\mu }J^{0}+\omega
\delta _{k}^{\mu }J^{k}$ is the conserved current of the model and
$J^{\mu }=i[\phi (D^{\mu }\phi )^{\ast }-\phi ^{\ast }(D^{\mu }\phi )]$\
is the conventional current density. Similarly, the equation of
motion of the Higgs field is
\begin{eqnarray}
0 &=&(\partial _{0}\omega _{1})D_{0}\phi +\omega _{1}D_{0}(D_{0}\phi )-\frac{%
\partial \omega _{1}}{\partial \phi ^{\ast }}|D_{0}\phi |^{2}  \label{geEq3}
\\[0.1in]
&&-(\partial _{k}\omega )D_{k}\phi -\omega D_{k}(D_{k}\phi )+\frac{\partial
\omega }{\partial \phi ^{\ast }}|D_{k}\phi |^{2}+\frac{\partial V}{\partial
\phi ^{\ast }}.  \notag
\end{eqnarray}

From Eq. (\ref{ge0}), the Gauss law reads
\begin{equation}
\kappa B=e\omega _{1}J_{0},  \label{geEq1}
\end{equation}%
we observe the Gauss law of Chern-Simons dynamics is modified by the
function $\omega _{1}(\left\vert \phi \right\vert )$ such that now the
conserved charge associated with the $U(1)$ global symmetry is given by
\begin{equation}
Q=\int \!d^{2}x\,e\omega _{1}J^{0}
\end{equation}%
however such as it happens in usual CSH model, the electric charge is
nonnull and proportional to the magnetic flux:
\begin{equation}
Q=\kappa \int \!d^{2}x~B=\kappa \Phi .
\end{equation}%
Therefore, independently the functional form of the generalizing functions $%
\omega _{1}(\left\vert \phi \right\vert )$ and $\omega (\left\vert \phi
\right\vert )$, the solutions always will be electrically charged.

Likewise, the Amp\`{e}re law reads%
\begin{equation}
\frac{\kappa }{2}\epsilon ^{k\alpha \beta }F_{\alpha \beta }+e\omega J_{k}=0.
\label{geEq2}
\end{equation}

Along the remain of the manuscript, we are interested in time-independent
soliton solutions that ensure the finiteness of the action (\ref{Acg1}).
These are the stationary points of the energy which for the static \textbf{\
\ }
\begin{equation}
E=\!\!\int \!d^{2}x\!\left[ -\kappa A_{0}B-e^{2}\omega _{1}A_{0}^{2}|\phi
|^{2}+\omega |D_{i}\phi |^{2}+V(|\phi |)\right] .  \label{EJP}
\end{equation}%
From the statitic Gauss law, we obtain the relation
\begin{equation}
A_{0}=-\frac{\kappa }{2e}\frac{B}{\omega _{1}|\phi |^{2}},  \label{A0}
\end{equation}%
which substituted in Eq. (\ref{EJP}) leads to the following expression for
the energy:
\begin{equation}
E=\int \!d^{2}x\left[ \frac{\kappa ^{2}}{4e^{2}}\frac{B^{2}}{\omega
_{1}|\phi |^{2}}+\omega |D_{i}\phi |^{2}+V(|\phi |)\right] .  \label{EJP1}
\end{equation}%
To proceed, we need the fundamental identity
\begin{equation}
|D_{i}\phi |^{2}=|D_{\pm }\phi |^{2}\pm eB|\phi |^{2}\pm \frac{1}{2}\epsilon
_{ik}\partial _{i}J_{k}  \label{iden}
\end{equation}%
where $D_{\pm }\phi =D_{1}\phi \pm iD_{2}\phi $. Then, by using (\ref{iden}%
), we may rewrite the energy (\ref{EJP1}) as
\begin{eqnarray}
E &=&\int \!d^{2}x\left[ \frac{\kappa ^{2}}{4e^{2}}\frac{B^{2}}{\omega
_{1}|\phi |^{2}}+V(|\phi |)+\omega |D_{\pm }\phi |^{2}\right.  \notag \\
&&\hspace{1.25cm}\left. \pm e\omega B\left\vert \phi \right\vert ^{2}\pm
\frac{1}{2}\omega \epsilon _{ik}\partial _{i}J_{k}\right] .  \label{EJP2}
\end{eqnarray}

We observe that the function $\omega (|\phi |)$ in the term\ $\omega
\epsilon _{ik}\partial _{i}J_{k}$ preclude us to implement the  BPS
procedure, i.e, the integrand must be expressed like a sum of squared terms
plus a total derivative plus a term proportional to the magnetic
field. Therefore, the key question is about the functional form of $\omega
(|\phi |)$ allowing a well defined implementation of the BPS formalism. We
start the searching of the function $\omega (|\phi |)$ from the following
expression:
\begin{equation}
\epsilon _{ik}\partial _{i}(\omega J_{k})=\omega \epsilon _{ik}\partial
_{i}J_{k}+\epsilon _{ik}(\partial _{i}\omega )J_{k}.  \label{eqw1}
\end{equation}%
By manipulating the last term $\epsilon _{ik}(\partial _{i}\omega )J_{k}$ it
reads%
\begin{equation}
\epsilon _{ik}(\partial _{i}\omega )J_{k}=\frac{\partial \omega }{\partial
\left\vert \phi \right\vert ^{2}}\left( \partial _{i}\left\vert \phi
\right\vert ^{2}\right) \epsilon _{ik}J_{k},  \label{eqw2}
\end{equation}%
where we have used the fact of $\omega $ be a explicit function of $|\phi
|^{2}$. After some algebra the term $\epsilon _{ik}\left( \partial
_{i}\omega \right) J_{k}$ becomes%
\begin{equation}
\epsilon _{ik}(\partial _{i}\omega )J_{k}=\left\vert \phi \right\vert ^{2}%
\frac{\partial \omega }{\partial \left\vert \phi \right\vert ^{2}}\epsilon
_{ik}\partial _{i}J_{k}+2eB\left\vert \phi \right\vert ^{4}\frac{\partial
\omega }{\partial \left\vert \phi \right\vert ^{2}}.  \label{eqw3}
\end{equation}%
Substituting this equation in (\ref{eqw1}) we arrive to
\begin{equation}
\epsilon _{ik}\partial _{i}(\omega J_{k})=\left( \omega +|\phi |^{2}\frac{%
\partial \omega }{\partial |\phi |^{2}}\right) \epsilon _{ik}\partial
_{i}J_{k}+2eB|\phi |^{4}\frac{\partial \omega }{\partial |\phi |^{2}}.
\label{eqw4}
\end{equation}

Here we impose that the function $\omega $ satisfies the following equation:
\begin{equation}
\omega +\left\vert \phi \right\vert ^{2}\frac{\partial \omega }{\partial
\left\vert \phi \right\vert ^{2}}=\beta \omega ,  \label{eqw5}
\end{equation}%
with $\beta$ a real constant. By solving Eq. (\ref{eqw5}) we obtain
the explicit functional form of $\omega (|\phi |)$,
\begin{equation}
\omega =C|\phi |^{2\beta -2},  \label{omegaa}
\end{equation}%
where the constant $C$ adjusts conveniently the mass dimension of $\omega $.

The key condition (\ref{eqw5}) allows to rewrite Eq. (\ref{eqw4}) in a more
simplified form
\begin{equation}
\epsilon _{ik}\partial _{i}(\omega J_{k})=\beta \omega \epsilon
_{ik}\partial _{i}J_{k}+2e(\beta -1)\omega B|\phi |^{2},  \label{eqw6}
\end{equation}%
allowing to write the term $\omega \epsilon _{ik}\partial _{i}J_{k}$
in the following way
\begin{equation}
\omega \epsilon _{ik}\partial _{i}J_{k}=\frac{1}{\beta }\epsilon
_{ik}\partial _{i}\left( \omega J_{k}\right) -2e\frac{\beta -1}{\beta }%
\omega B\left\vert \phi \right\vert ^{2}.  \label{eqw7}
\end{equation}%
By introducing it in Eq. (\ref{EJP2}), the energy becomes
\begin{eqnarray}
E &=&\!\int \!\!d^{2}x\left[ \frac{\kappa ^{2}}{4e^{2}}\frac{B^{2}}{\omega
_{1}|\phi |^{2}}+V(|\phi |)+\omega |D_{\pm }\phi |^{2}\right.  \label{EJP3}
\\
&&\hspace{1.25cm}\left. \pm \frac{1}{\beta }e\omega B\left\vert \phi
\right\vert ^{2}\pm \frac{1}{2\beta }\epsilon _{ik}\partial _{i}\left(
\omega J_{k}\right) \right] .  \notag
\end{eqnarray}%
We write the two first terms as
\begin{eqnarray}
\frac{\kappa ^{2}}{4e^{2}}\frac{B^{2}}{|\phi |^{2}\omega _{1}}+V &=&\frac{%
\kappa ^{2}}{4e^{2}}\frac{1}{|\phi |^{2}\omega _{1}}\left( B\mp \frac{2e}{%
\kappa }\left\vert \phi \right\vert \sqrt{\omega _{1}V}\right) ^{2}  \notag
\\
&&\pm \frac{\kappa B}{e|\phi |}\sqrt{\frac{V}{\omega _{1}}}
\end{eqnarray}%
By substituting in (\ref{EJP3}), we have
\begin{eqnarray}
E &=&\!\!\int \!\!d^{2}x\left[ \frac{\kappa ^{2}}{4e^{2}}\frac{1}{|\phi
|^{2}\omega _{1}}\left( B\mp \frac{2e\left\vert \phi \right\vert }{\kappa }
\sqrt{\omega _{1}V}\right) ^{2}\right.  \notag \\
&&\hspace{1cm}\left. +\omega |D_{\pm }\phi |^{2}\pm \frac{1}{2\beta }
\epsilon _{ik}\partial _{i}\left( \omega J_{k}\right) \right.  \notag \\
&&\hspace{1cm}\left. \pm B\left( \frac{\kappa }{e|\phi |}\sqrt{\frac{V}{
\omega _{1}}}+\frac{1}{\beta }e\omega \left\vert \phi \right\vert
^{2}\right) \right] .  \label{EPJ4}
\end{eqnarray}%
To finish the BPS procedure, we observe that if in the third row the
term multiplying to the magnetic field is equal to $ev^{2}$, it
allows to define explicitly the form of the potential $V(|\phi |)$,
\begin{equation}
V\left( |\phi |\right) =\frac{e^{4}v^{4}}{\kappa ^{2}}\omega _{1}|\phi
|^{2}\left( 1-\frac{\left\vert \phi \right\vert ^{2\beta }}{v^{2\beta }}
\right) ^{2},  \label{potsbps0}
\end{equation}%
where we have substituted the explicit form of $\omega \left( |\phi
|\right) $ given by Eq. (\ref{omegaa})  with  $C=\beta v^{2-2\beta}$
in order to the vacuum expectation value of the Higgs field to be
$|\phi |=v$. The function $\omega _{1}\left( |\phi |\right) $
still remains arbitrary. Hence, the energy (\ref{EPJ4}) reads
\begin{eqnarray}
E &&\!\!=\!\!\int \!\!d^{2}x\left\{ \pm ev^{2}B\pm \frac{1}{2\beta }\epsilon
_{ik}\partial _{i}\left( \omega J_{k}\right) +\omega |D_{\pm }\phi
|^{2}\right. \\
&&\left. +\frac{\kappa ^{2}}{4e^{2}|\phi |^{2}\omega _{1}}\left[ B\mp \frac{%
2e^{3}v^{2}}{\kappa ^{2}}\omega _{1}|\phi |^{2}\left( 1-\frac{\left\vert
\phi \right\vert ^{2\beta }}{v^{2\beta }}\right) \right] ^{2}\right\} .
\notag
\end{eqnarray}%
We see that under appropriated boundary conditions the total derivative
gives null contribution to the energy. Then, the energy is bounded below by
a multiple of the magnetic flux magnitude (for positive flux we choose the
upper signs, and for negative flux we choose the lower signs):
\begin{equation}
E\geq \pm ev^{2}\!\!\int \!\!d^{2}xB=ev^{2}|\Phi |.
\end{equation}%
This bound is saturated by fields satisfying the Bogomolnyi or
self-dual equations \cite{Bogo}
\begin{equation}
D_{\pm }\phi =0,  \label{BPS1}
\end{equation}%
\begin{equation}
B=\pm \frac{2e^{3}v^{2}}{\kappa ^{2}}\omega _{1}|\phi |^{2}\left( 1-\frac{%
\left\vert \phi \right\vert ^{2\beta }}{v^{2\beta }}\right) .  \label{BPS20}
\end{equation}%
If we require that the magnetic field be nonsingular at origin, the function
$\omega _{1}(|\phi |)$ should behave like $|\phi |^{2\delta }$ with $\delta
\geq -1$. On the other hand, positivity and finiteness of the BPS energy
density requires $\beta \geq 1$.

Below we study interesting models by given a specific form of the functions $%
\omega $ and $\omega _{1}$.

\section{Some simple models}

In the following we analyze some interesting but simple models by setting,
\begin{equation}
\omega (|\phi |)=\left( N+1\right) \frac{|\phi |^{2N}}{v^{2N}}~,~\ \omega
_{1}(|\phi |)=\frac{|\phi |^{2M}}{v^{2M}}.  \label{omega}
\end{equation}
The BPS potential (\ref{potsbps0}) reads
\begin{equation}
V\left( |\phi |\right) =\frac{e^{4}v^{6}}{\kappa ^{2}}\frac{|\phi |^{2M+2}}{%
v^{2M+2}}\left( 1-\frac{\left\vert \phi \right\vert ^{2N+2}}{v^{2N+2}}%
\right) ^{2},  \label{potsbps}
\end{equation}%
and the BPS equation (\ref{BPS20}) becomes
\begin{equation}
B=\pm \frac{2e^{3}v^{4}}{\kappa ^{2}}\frac{|\phi |^{2M+2}}{v^{2M+2}}\left( 1-%
\frac{\left\vert \phi \right\vert ^{2N+2}}{v^{2N+2}}\right) .  \label{BPS2}
\end{equation}

Here, it interesting to note that for $N=0$ and  $M=0$ the self-duality
equations (\ref{BPS1}) and (\ref{BPS2}), becomes the well known Bogomolnyi
equations of the Chern-Simons-Higgs theory \cite{JW1}, \cite{JW2},
\begin{equation}
D_{\pm }\phi =0,\ ~B=\pm \frac{2e^{3}}{\kappa ^{2}}|\phi |^{2}\left(
v^{2}-|\phi |^{2}\right) .
\end{equation}
In the case $N=0$ and  $M=-1$, we have
\begin{equation}
D_{\pm }\phi =0,\ ~B=\pm \frac{2e^{3}v^{2}}{\kappa ^{2}}\left( v^{2}-|\phi
|^{2}\right) .  \label{BPS11}
\end{equation}%
These equations are essentially the Bogomolnyi equations of the
Maxwell-Higgs model, whose solutions are the well known Nielsen-Olesen
vortices \cite{NO}. The difference lies in the fact that, here, our
self-dual solitons not only carry magnetic flux, as in the Higgs model, but
also $U(1)$ charge. This is a consequence that in our theory the dynamics of
gauge field is dictated by a Chern-Simons term instead of a Maxwell term as
in Maxwell-Higgs theory. So, for $N=0$ and $M=-1$, we obtain self-dual
configurations which are mathematically identical to the Nielsen-Olesen ones
but differently our solutions have electric charge.

\subsection{Vortex configurations}

Specifically, we look for axially symmetric solutions using the standard
static vortex \emph{Ansatz}
\begin{equation}
\phi =vg(r)e^{in\theta },~A_{\theta }=-\frac{a(r)-n}{er}
\end{equation}

The \emph{Ansatz} allows to express the magnetic field as%
\begin{equation}
B=-\frac{a^{\prime }}{er}
\end{equation}%
where $\,^{\prime }\,$ denotes a derivative in relation to the coordinate $r$%
. Likewise, the BPS equations (\ref{BPS1}) and (\ref{BPS2}) are written as%
\begin{equation}
g^{\prime }=\pm \frac{ag}{r},  \label{BPS1_1}
\end{equation}%
\begin{equation}
B=-\frac{a^{\prime }}{er}=\pm \frac{2e^{3}v^{4}}{\kappa ^{2}}g^{2M+2}\left(
1-g^{2N+2}\right) .  \label{BPS2_1}
\end{equation}%
These equations are solved considering the profiles $g$ and $a$ are well
behaved functions satisfying the following boundary conditions
\begin{eqnarray}
g(0) &=&0,\;a(0)=n,  \label{b} \\[0.2cm]
g(\infty ) &=&1,\,a(\infty )=0.  \label{b1}
\end{eqnarray}

The BPS energy density of the model reading from%
\begin{equation}
E_{_{BPS}}=2\pi \int \!dr~r\varepsilon _{_{BPS}},
\end{equation}%
is given by
\begin{eqnarray}
\varepsilon _{_{BPS}}&=&\frac{2e^{4}v^{6}}{\kappa ^{2}}g^{2M+2}\left(
1-g^{2N+2}\right) ^{2}\nonumber\\
&&+2v^{2}\left( N+1\right) g^{2N}\left( \frac{ag}{r}\right) ^{2},
\end{eqnarray}
the requirement of finite energy density, for all values of the winding
number $n$, imposes $N\geq 0$ and $M\geq -1$.

\subsection{Checking the boundary conditions}

We obtain the behavior of the solutions of Eqs. (\ref{BPS1_1}) and (\ref%
{BPS2_1}) in the neighborhood of $r\rightarrow 0$ using power series method,
\begin{eqnarray}
g(r) &=&G_{n}r^{n}-\frac{e^{4}v^{4}\left( G_{n}\right) ^{2M+3}r^{n\left(
2M+3\right) +2}}{2\kappa ^{2}\left( nM+n+1\right) ^{2}}+{\mathcal{\ldots }}
\quad  \label{bc0_g1} \\[0.08in]
a(r) &=&n-\frac{{}}{{}}\frac{e^{4}v^{4}\left( G_{n}\right) ^{2M+2}r^{n\left(
2M+2\right) +2}}{\kappa ^{2}\left( nM+n+1\right) }+{\mathcal{\ldots }}.
\label{bc0_a}
\end{eqnarray}%
It verifies the boundary conditions given in Eq. (\ref{b}).

For $r\rightarrow +\infty $, the behavior of the soliton solutions becomes
similar to the Nielsen-Olesen vortices,
\begin{eqnarray}
g(r) &\sim &1-G_{_{\infty }}\frac{e^{-m_{s}r}}{\sqrt{r}}~  \label{inf_1} \\%
[0.15cm]
a(r) &\sim &G_{_{\infty }}m_{s}\sqrt{r}\,e^{-m_{s}r},
\end{eqnarray}%
where $G_{_{\infty }}$ is a numerical constant determined numerically and $%
m_{s}$, the self-dual mass of the bosonic fields, is given by
\begin{equation}
m_{s}=\frac{2e^{2}v^{2}}{\kappa }\sqrt{N+1}.  \label{beta}
\end{equation}%
It is verified that for $N=0$, the mass scale is exactly the one of the
Chern-Simons-Higgs model.%
\begin{equation}
V\left( |\phi |\right) =\frac{e^{4}v^{6}}{\kappa ^{2}}\frac{|\phi |^{2M+2}}{%
v^{2M+2}}\left( 1-\frac{\left\vert \phi \right\vert ^{2N+2}}{v^{2N+2}}%
\right) ^{2},
\end{equation}

\subsection{Numerical analysis}

Below, without loss of generality we set $e=1$, $v=1$, $\kappa =1$.

Before performing the numerical solution of the self-dual equations (\ref%
{BPS1_1}) and (\ref{BPS2_1}) we do the following observations in relation to
the BPS potential (\ref{potsbps}): First, it provides a $|\phi |^{4}$
potential for $M=-1$ and $N=0$,
\begin{equation}
V(|\phi |)=\left( 1-|\phi |^{2}\right) ^{2},  \label{phi4}
\end{equation}

Second, the BPS potential also provides a family of $|\phi |^{6}$ potentials
when the condition $M=-2N$ is satisfied and $N$ is restricted to the
interval $0\leq N\leq 1/2$,
\begin{equation}
V(|\phi |)=|\phi |^{-4N+2}\left( 1-|\phi |^{2N+2}\right) ^{2}.
\end{equation}

Below, our numerical analysis considers only these two potentials to solve
the BPS equations (\ref{BPS1_1}) and (\ref{BPS2_1}). In particular, we solve
the Bogomolnyi equations only for winding number $n=1$.

In the figures, the red line represents the case $M=-1$ and $N=0$ providing
the Nielsen-Olesen-like vortices, whereas the blue lines depict the vortex
solutions for the values of $M$ and $N$ generating some $|\phi|^{6}$%
-potentials. In particular, we have plotted three solutions in blue lines:

\begin{itemize}
\item $M=0$\ and $N=0$, which generates the well know Chern-Simons-Higgs
vortices.

\item $M=-0.5$\ and $N=0.25$, associated to the self-dual potential
\begin{equation}
V(|\phi |)=|\phi |\left( 1-|\phi |^{\frac{5}{2}}\right) ^{2}.  \label{ffs}
\end{equation}

\item $M=-1$ and $N=0.5$, associated to the self-dual potential
\begin{equation}
V(|\phi |)=\left( 1-|\phi |^{3}\right) ^{2}.  \label{phi6}
\end{equation}
\end{itemize}

\begin{figure}[]
\centering
\includegraphics[width=8.5cm]{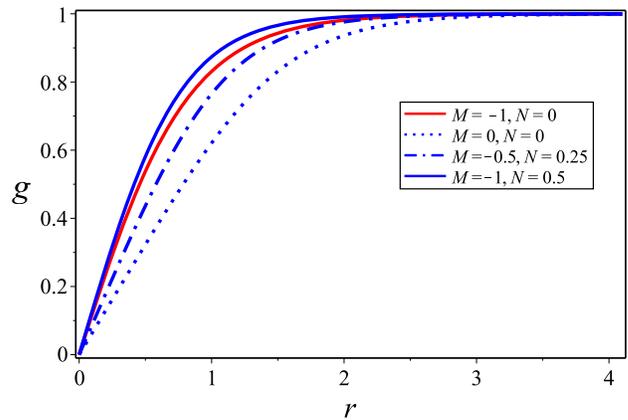}
\caption{ The profiles of the Higgs field $g(r)$ for $n=1$. The red lines
represent the solutions for a $|\protect\phi|^4$ potential and blue lines
for $|\protect\phi|^6$ potentials.}
\label{f_higgs}
\end{figure}

\begin{figure}[]
\centering
\includegraphics[width=8.5cm]{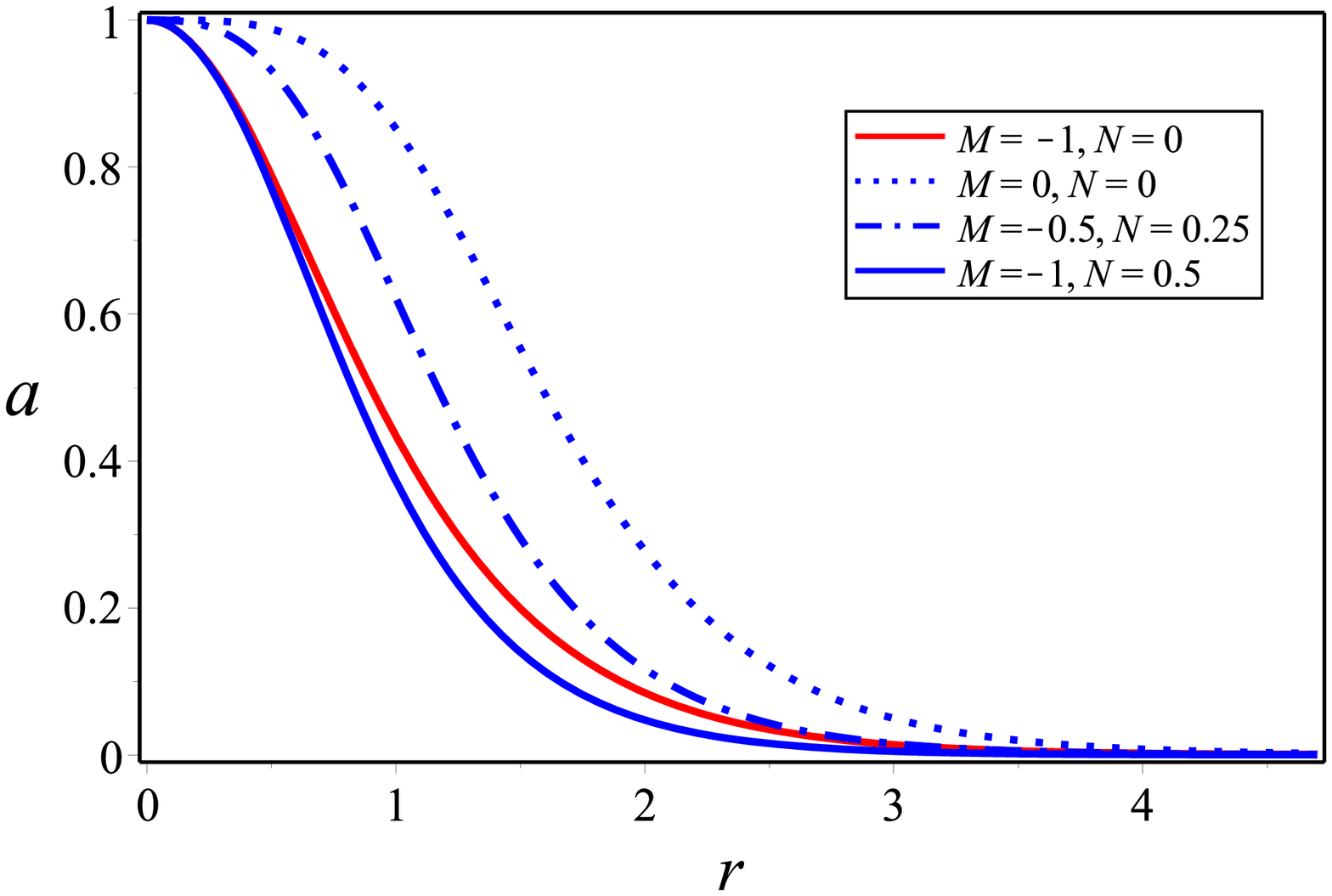}
\caption{ The profiles of the gauge field $a(r)$ for $n=1$. The red lines
represent the solutions for a $|\protect\phi|^4$ potential and blue lines
for $|\protect\phi|^6$ potentials.}
\label{f_gauge}
\end{figure}

\begin{figure}[]
\centering
\includegraphics[width=8.5cm]{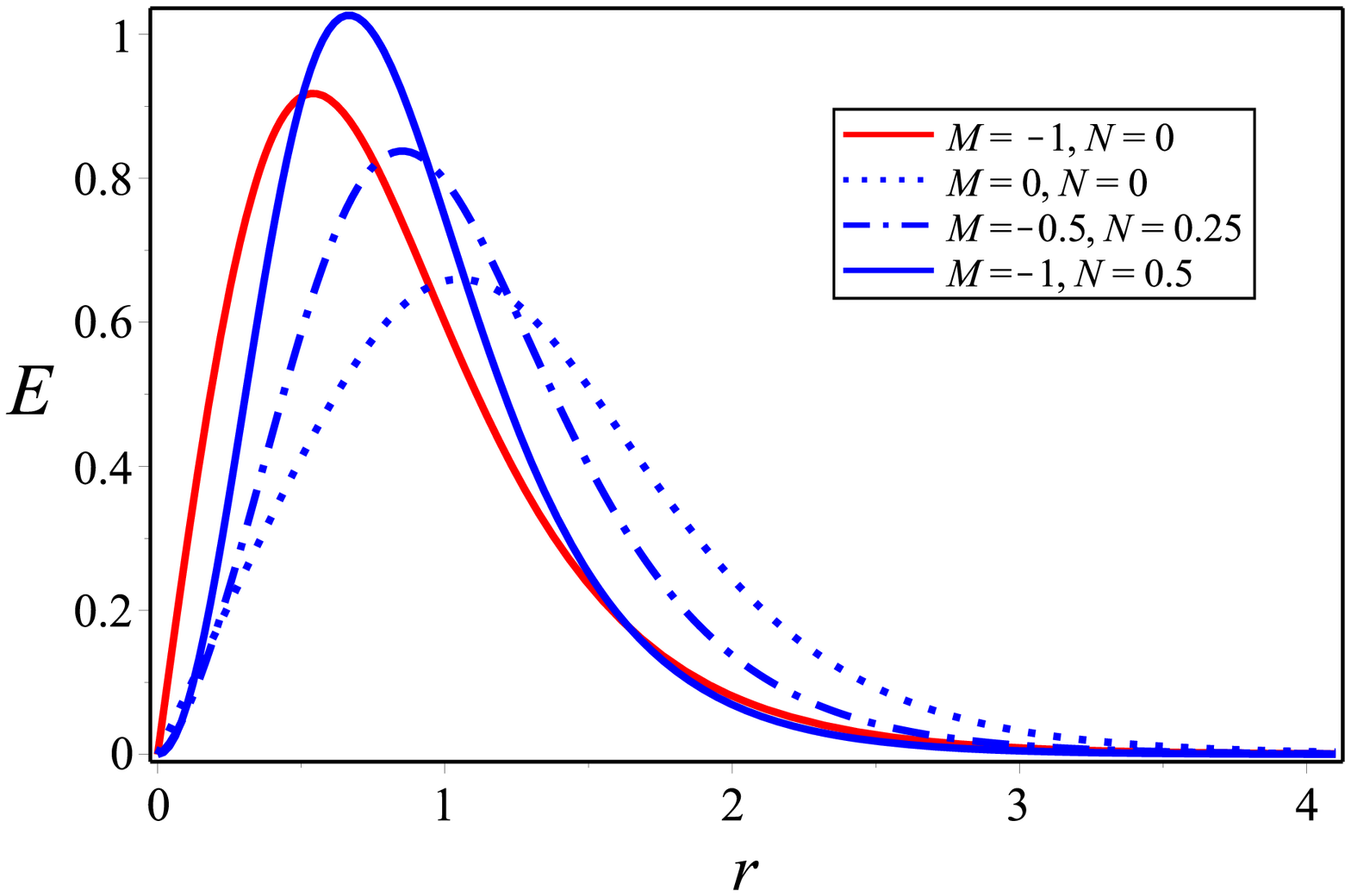}
\caption{ The profiles of the electric field $E(r)=A^{\prime }_0(r)$ for $%
n=1 $. The red lines represent the solutions for a $|\protect\phi|^4$
potential and blue lines for $|\protect\phi|^6$ potentials.}
\label{f_electric}
\end{figure}

\begin{figure}[]
\centering
\includegraphics[width=8.5cm]{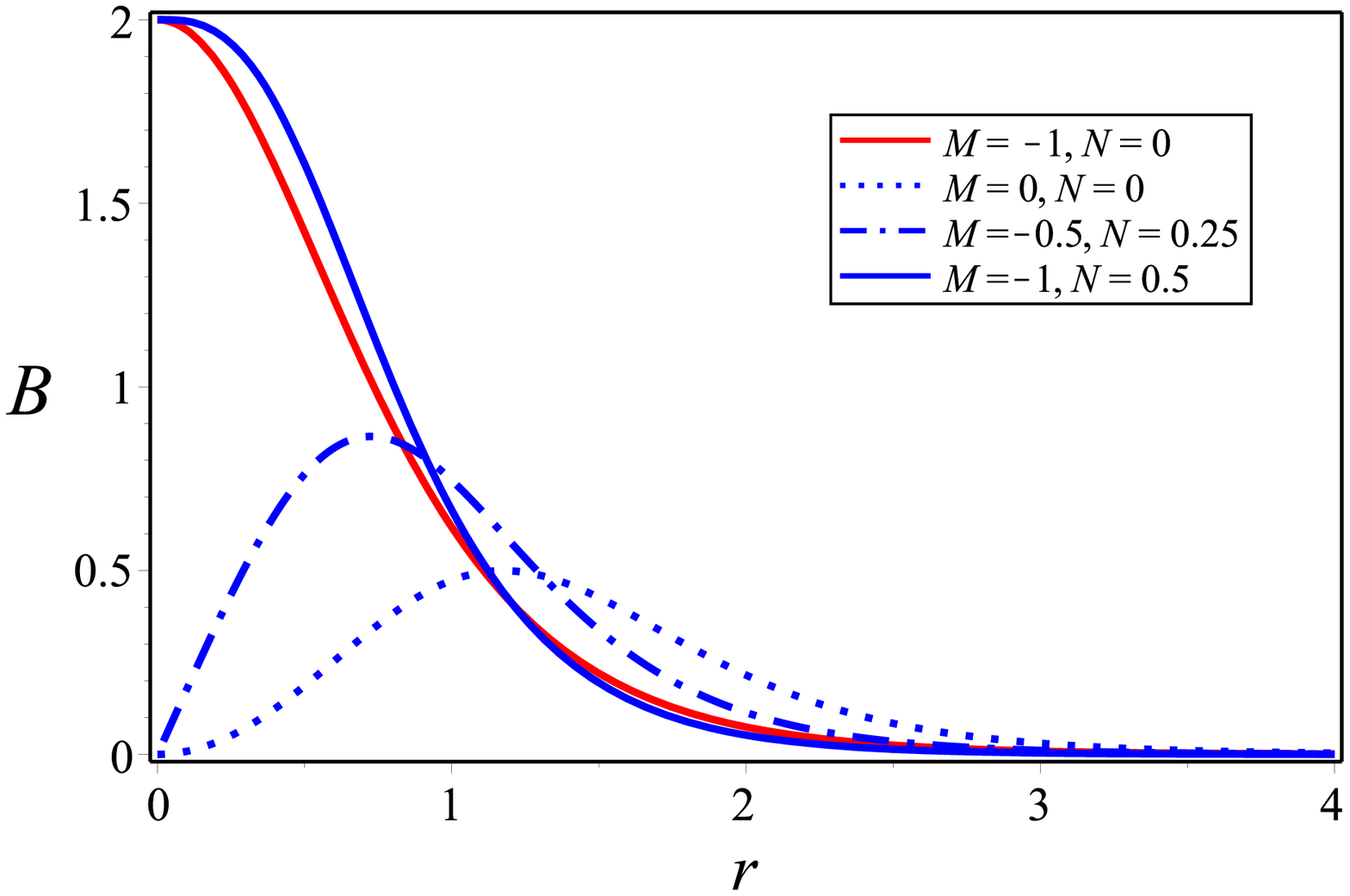}
\caption{ The profiles of the magnetic field $B(r)$ for $n=1$. The red lines
represent the solutions for a $|\protect\phi|^4$ potential and blue lines
for $|\protect\phi|^6$ potentials.}
\label{f_magnetic}
\end{figure}

Note that in the cases $M=-1$, $N=0$ and $M=-1$, $N=0.5$, i.e. the cases
associated to the potentials (\ref{phi4}) and (\ref{phi6}), there is only
one degenerate vacua at $|\phi |=1$. This fact leads us to similar
solutions, which can be appreciated in Figs. \ref{f_higgs}, \ref{f_gauge}, %
\ref{f_magnetic}, \ref{f_electric}.

\begin{figure}[]
\centering
\includegraphics[width=8.5cm]{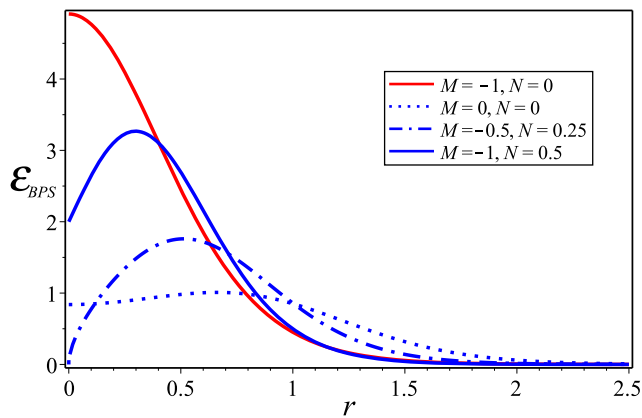}
\caption{ The profiles of the BPS energy density $\protect\varepsilon%
_{_{bps}} (r)$ for $n=1$. The red lines represent the solutions for a $|%
\protect\phi|^4$ potential and blue lines for $|\protect\phi|^6$ potentials.}
\label{f_energia}
\end{figure}

For the cases where $M\neq -1$, the $|\phi|^6$ potential have two vacua: $%
|\phi|=0$ and $|\phi|=1$. In these cases, the profiles of the magnetic field
are rings whose maximum amplitude, for increasing values of $N$, approaches
to the origin (see Fig. \ref{f_magnetic}). Also the profiles of the BPS
energy density have a ring-like format ((see Fig. \ref{f_energia}) and the
ring format is explicit for $n>1$.

On the other hand, for electric field, whenever the values of $M$ and $N$
here considered, the profiles always are rings around the origin (see Fig. %
\ref{f_electric}).

\section{Remarks and conclusions}

In summary, we have proposed a generalized abelian Chern-Simons-Higgs model
with explicit breaking of Lorentz covariance and explored the respective
Bogomolnyi framework. During the implementation of the BPS trick it is shown
that the generalized functions should satisfy some requirements: The
function $\omega (|\phi |)$ must be a monomial, i.e., $\omega =C|\phi
|^{2\beta -2}$ for all $\beta \geq 1$ and the function $\omega _{1}(|\phi |)$
must be regular at the origin ($\omega _{1}\propto |\phi |^{2\delta }$ with
$\delta \geq -1)$.  Under such conditions imposed on the generalized
functions,  it is guaranteed the existence of self-dual solitonic configurations
satisfying Bogomolnyi equations whose  magnetic field and  BPS energy density
are well behaved. As we expected, the infinity family of self-dual configurations
have finite  energy which is proportional to the magnitude of the magnetic flux.
In particular, we have studied the
self-dual vortices provided by the choice $\omega _{1}(|\phi |)=\left(
N+1\right) ~|\phi |^{2N}$ and $\omega _{1}(|\phi |)=|\phi |^{2M}$. It was
shown the vortex solutions of the Maxwell-Higgs model and the
Chern-Simons-Higgs model can be also obtained. Besides that, we have constructed
two new solitonic solutions which correspond to Chern-Simons theory coupled
to two types of $|\phi |^{6}$ potentials given by Eqs. (\ref{ffs})  and
(\ref{phi6}), respectively.

Finally, it is worthwhile to point out that existence of BPS states is
linked to the existence of a $\mathcal{N}=2$-extended supersymmetric model
\cite{WittenOlive}. We are studying such a possibility despite the fact that
in this model the Lorentz symmetry is explicitly broken. Advances in this
direction will be reported elsewhere.

\subsection*{Conflict of Interests}

The authors declare that there is no conflict of interests regarding the
publication of this paper.

\begin{acknowledgments}
R.C. thanks to CNPq, CAPES and FAPEMA (Brazilian agencies) by financial
support. L.S. is supported by CONICET.
\end{acknowledgments}


\begin{thebibliography}{99}
\bibitem{Jk1} S. K. Paul, A. Khare, Phys. Lett. B\textbf{174}, 420 (1986)
[Erratum-ibid. 177B, 453 (1986)].

\bibitem{Jk2} H. J. de Vega, F. A. Schaposnik, Phys. Rev. D\textbf{34}, 3206
(1986).

\bibitem{hor1} R. Jackiw and S. Y. Pi, Prog. Theor. Phys. Suppl. \textbf{107}%
, 1 (1992).

\bibitem{hor2} G. V. Dunne, \emph{Self-Dual Chern–Simons Theories}, Lecture Notes in Physics, m36, 1995, Springer.

\bibitem{hor3} G. V. Dunne, [arXiv:hep-th/9902115].

\bibitem{hor4} F. A. Schaposnik, [arXiv:hep-th/0611028].

\bibitem{hor5} P. A. Horvathy, P. Zhang, Phys. Rept. \textbf{481}, 83 (2009).

\bibitem{JW1} R. Jackiw, E. J. Weinberg, Phys. Rev. Lett. \textbf{64}, 2234
(1990).

\bibitem{JW2} J. Hong, Y. Kim, P. Y. Pac, Phys. Rev. Lett. \textbf{64}, 2230
(1990).

\bibitem{JP1} R. Jackiw, S. Y. Pi, Phys. Rev. Lett. \textbf{64}, 2969 (1990).

\bibitem{JP2} R. Jackiw, S. Y. Pi, Phys. Rev. D \textbf{42}, 3500 (1990);
Erratum-ibid. D \textbf{48}, 3929 (1993).

\bibitem{Bogo} E. Bogomolyi, Sov. J. Nucl. Phys \textbf{24}, 449 (1976).

\bibitem{Vega} H. de Vega, F .A. Schaposnik, Phys. Rev. D\textbf{14}, 1100
(1976).

\bibitem{LLW} C. Lee, K. Lee, E. J. Weinberg, Phys. Lett. B\textbf{243}, 105
(1990).

\bibitem{JLW} R. Jackiw, Ki-Myeong Lee, E. J. Weinberg, Phys. Rev. D\textbf{%
42}, 3488 (1990).

\bibitem{sy} T. H. R. Skyrme, Proc. Roy. Soc. A\textbf{262}, 237 (1961).

\bibitem{APDM1} C. Armendariz-Picon, T. Damour and V. Mukhanov, Phys. Lett. B%
\textbf{458}, 209 (1999).

\bibitem{APDM2} C. Armendariz-Picon, V. Mukhanov, P. J. Steinhardt, Phys.
Rev. Lett. \textbf{85}, 4438 (2000).

\bibitem{APDM3} C. Armendariz-Picon, V. Mukhanov, Paul J. Steinhardt, Phys.
Rev. D\textbf{63}, 103510 (2001).

\bibitem{APDM4} T. Chiba, T. Okabe, M. Yamaguchi, Phys. Rev. D\textbf{62},
023511 (2000).

\bibitem{APDM5} M. Malquarti, E. J. Copeland, A. R. Liddle, Phys. Rev. D%
\textbf{68}, 023512 (2003).

\bibitem{APDM6} J. U. Kang, V. Vanchurin, S. Winitzki, Phys. Rev. D\textbf{76%
}, 083511 (2007).

\bibitem{APDM7} E. Babichev, V. Mukhanov, A. Vikman, J. High Energy Phys.
\textbf{02}, 101 (2008).

\bibitem{12} A. Sen, JHEP 0207, 065 (2002).

\bibitem{131} N. Arkani-Hamed, H.-C. Cheng, M. A. Luty, S. Mukohyama, JHEP
\textbf{0405}, 074 (2004).

\bibitem{132} N. Arkani-Hamed, P. Creminelli, S. Mukohyama, M. Zaldarriaga,
JCAP \textbf{0404}, 001 (2004).

\bibitem{133} S. Dubovsky, JCAP \textbf{0407}, 009 (2004).

\bibitem{134} D. Krotov, C. Rebbi, V. Rubakov, V. Zakharov, Phys.Rev. D%
\textbf{71}, 045014 (2005).

\bibitem{135} A. Anisimov, A. Vikman, JCAP \textbf{0504}, 009 (2005).

\bibitem{MV} V. Mukhanov and A. Vikman, J. Cosmol. Astropart. Phys. \textbf{%
02}, 004 (2006).

\bibitem{APL} C. Armendariz-Picon and E. A. Lim, J. Cosmol. Astropart. Phys.
\textbf{08}, 007 (2005).

\bibitem{BAi1} D. Bazeia, E. da Hora, C. dos Santos, R. Menezes, Phys. Rev. D%
\textbf{81}, 125014 (2010).

\bibitem{BAi2} D. Bazeia, E. da Hora, R. Menezes, H. P. de Oliveira, C. dos
Santos, Phys. Rev. D\textbf{81}, 125016 (2010).

\bibitem{BAi3} C. dos Santos, E. da Hora, Eur. Phys. J. C\textbf{70}, 1145
(2010);

\bibitem{BAi4} C. dos Santos, E. da Hora, Eur. Phys. J. C\textbf{71}, 1519
(2011).

\bibitem{BAi5} C. dos Santos, Phys. Rev. D\textbf{82}, 125009 (2010).

\bibitem{BAi6} D. Bazeia, E. da Hora, C. dos Santos, R. Menezes, Eur. Phys.
J. C\textbf{71}, 1833 (2011).

\bibitem{BAi7} D. Bazeia, R. Casana, E. da Hora, R. Menezes, Phys. Rev. D%
\textbf{85}, 125028 (2012).

\bibitem{BAi8} R. Casana, M. M. Ferreira, Jr., E. da Hora, Phys. Rev. D%
\textbf{86} 085034 (2012).

\bibitem{SG1} E. Babichev, Phys. Rev. D\textbf{74}, 085004 (2006).

\bibitem{SG2} E. Babichev, Phys. Rev. D\textbf{77}, 065021 (2008).

\bibitem{SG3} C. Adam, J. Sanchez-Guillen, A. Wereszczynski, J. Phys. A
\textbf{40}, 13625 (2007); Erratum-ibid. \textbf{42}, 089801 (2009).

\bibitem{SG4} C. Adam, N. Grandi, J. Sanchez-Guillen, A. Wereszczynski, J.
Phys. A\textbf{41}, 212004 (2008); Erratum- ibid. \textbf{42}, 159801 (2009).

\bibitem{SG5} C. Adam, N. Grandi, P. Klimas, J. Sanchez-Guillen, A.
Wereszczynski, J. Phys. A\textbf{41}, 375401 (2008).

\bibitem{SG6} C. Adam, P. Klimas, J. Sanchez-Guillen, A. Wereszczynski, J.
Phys. A\textbf{42}, 135401 (2009).

\bibitem{lucas1} Lucas Sourrouille, Mod. Phys. Lett. A\textbf{30}, 1501211
(2015).

\bibitem{lucas2} Rodolfo Casana, Lucas Sourrouille, Mod. Phys. Lett. A%
\textbf{29}, 1450124 (2014).

\bibitem{lucas3} Lucas Sourrouille, Phy. Rev. D\textbf{87}, 067701 (2013).

\bibitem{lucas4} Lucas Sourrouille, Phy. Rev. D\textbf{86}, 085014 (2012).

\bibitem{B1} E. Babichev, Phys. Rev. D\textbf{74}, 085004 (2006).

\bibitem{B2} D. Bazeia, L. Losano, R. Menezes, J. C. R. E. Oliveira, Eur.
Phys. J. C\textbf{51}, 953 (2007).

\bibitem{B3} X. Jin, X. Li. and D. Liu, Classical Quantum Gravity \textbf{24}%
, 2773 (2007).

\bibitem{NO} H. B. Nielsen and P. Olesen, Nucl. Phys. B\textbf{61} 45 (1973).

\bibitem{WittenOlive} E. Witten and D. Olive, Phys. Lett. B\textbf{78}, 97
(1978).
\end{thebibliography}
\end{document}